\documentclass{optica-article}

\usepackage{lineno}

\begin{document}

\title{Low-coherence interferometry with undetected mid-infrared photons in the high-gain regime}

\author{Giovanni Zotti,\authormark{1,2,3,$\dagger$,*} 
Dmitri B. Horoshko,\authormark{4,5} 
Mikhail I. Kolobov,\authormark{5} 
Yoad Michael,\authormark{6} 
Ziv Gefen,\authormark{6} 
Maria V. Chekhova,\authormark{1,2,7} 
Kazuki Hashimoto\authormark{1,8,$\dagger$}}

\address{\authormark{1} Max Planck Institute for the Science of Light, Staudtstr. 2, 91058 Erlangen, Germany}
\address{\authormark{2} Friedrich-Alexander-Universität Erlangen-Nürnberg, Staudtstr. 7, 91058 Erlangen, Germany}
\address{\authormark{3} Department of Physics and Astronomy “A. Righi”, University of Bologna, I-40127 Bologna, Italy}
\address{\authormark{4} Institute for Quantum Optics, Ulm University, Albert-Einstein-Allee 11, 89081 Ulm, Germany}
\address{\authormark{5} Univ. Lille, CNRS, UMR 8523 - PhLAM - Physique des Lasers Atomes et Mol\'ecules, F-59000 Lille, France}
\address{\authormark{6} Raicol Crystals, Hamelacha 22, Rosh Ha’Ayin 4809162, Israel}
\address{\authormark{7} Faculty of Electrical and Computer Engineering, Technion—Israel Institute of Technology, Haifa 32000, Israel}
\address{\authormark{8} Institute for Photon Science and Technology, The University of Tokyo, Tokyo 113-0033, Japan}

\address{\authormark{$\dagger$} These authors contributed equally to this work.}
\address{\authormark{*} giovanni.zotti@mpl.mpg.de}

\begin{abstract}
We develop a high-parametric-gain SU(1,1) interferometer based on an aperiodically poled Potassium Titanyl Phosphate (apKTP) crystal, enabling frequency-domain low-coherence interferometry with undetected mid-infrared photons. The system achieves a signal-to-noise ratio as high as $40$ dB and axial resolution of $30$ $\mu$m, with a $3$ $\mu$m-centered idler beam. By increasing the poling-period range, we also improve the axial resolution to $17$ $\mu$m, demonstrating a straightforward route to enhance the performance by working on the crystal design.
\end{abstract}

Low-coherence interferometry (LCI) is a well-established, non-invasive optical technique that enables high-resolution axial imaging by inducing interference between a reference beam and the light backscattered or backreflected from a sample \cite{LCIandOCT}. It provides high sensitivity, micrometric axial resolution, and a typical imaging depth on the order of some $\mathrm{mm}$, making it particularly suitable for the investigation of fragile biological samples. Accordingly, its principal application lies in biomedical imaging, particularly through optical coherence tomography (OCT) \cite{drexler2008optical}. As a result, LCI systems are optimized to operate in the visible/near-infrared (NIR) spectral range, which also aligns with the availability of commercial light sources and detectors. \par
However, extending LCI into the mid-infrared (MIR) domain has gained growing interest due to its potential advantages in various fields, including quality control of manufactured ceramics \cite{Su_ceramics_1}, inspection of Ge or SiC based electronics \cite{SiC_OCT}, and the characterization of oil-based pigments for art conservation  \cite{Zorinpyroarray}.
Despite this promise, currently available MIR detectors are generally less efficient than their visible and NIR counterparts. Their performance is indeed strongly affected by the increased thermal noise, whose suppression requires expensive cooling systems. Attempts to implement LCI by employing such devices \cite{Zorinpyroarray, FirstMIROCT} have demonstrated severe limitations in terms of achievable sensitivity and cost.\par
 Upconversion-based MIR LCI, where the MIR probe is converted to visible/NIR light for detection, has also been successfully implemented, circumventing the need for MIR operating detectors \cite{Israelsen_upconversion,yagi_upconversion}. However, these implementations still necessitate high-power MIR sources, including MIR femtosecond or supercontinuum pulses, and an additional beam from another source or a filtered excitation beam for MIR source to drive the upconversion process. These requirements lead to increased complexity of the system.\par
 A different approach involves LCI with undetected photons, which relies on a quantum SU(1,1) interferometer architecture \cite{ChekhovaSU11} to circumvent, at once, the need for both MIR sources and detectors. In this scheme, a narrowband pump laser is directed into a nonlinear medium to generate broadband photon pairs at different wavelengths (signal and idler) via parametric down-conversion (PDC) or four-wave mixing (FWM). The idler (MIR) photons probe the sample, while the signal ones (visible or NIR) serve as a reference. Upon backreflection of pump, signal, and idler photons in the nonlinear medium, nonlinear interference occurs, leading to a signal spectrum analogous to a standard LCI one. Crucially, the full information about the sample can be retrieved by simply detecting the signal photons, enabling indirect probing via undetected MIR photons. Efficient implementation requires the generation of broadband pairs, typically achieved through quasi-phase-matching in engineered nonlinear media~\cite{TashimaQPM,VanselowQPM}. 
 Initial demonstrations of MIR LCI with undetected photons were presented in \cite{Paterova_first_LCI_und}. Subsequent work \cite{Vanselow:20} exploited a broadband PDC inside a periodically poled KTP crystal (ppKTP), achieving axial resolution down to
$10.1$ $\mu$m and signal-to-noise ratio of $69$ $\mathrm{dB}$ at $1$ s integration time, with only $90$ pW of MIR power probing the sample. All such implementations operated in the so-called low-gain regime of PDC, where the generated photons in each mode were so few that they did not stimulate the generation of new pairs, on both the first and the second pass through the nonlinear medium. This resulted in low signal intensity to be detected (order of $\mathrm{pW}$). \par
In contrast, the high-gain regime is characterized by exponential photon generation along the nonlinear medium, via optical parametric amplification (OPA). This process produces much higher signal powers at the output (from $\mathrm{nW}$ up to $\mu$W or even $\mathrm{mW}$  order \cite{Ishakov,mWtwinbeams}), accessible with ordinary photodetectors \cite{Horoshko2020,Machado}. Additionally, the use of a double-pass scheme combined with the stimulation effect allows one to non-invasively interrogate the sample with a weak idler and retrieve the information by detecting an intense signal \cite{Hashimotohighgainvis}. In this regime, the visibility scales nonlinearly with the the reflectivity of the sample, making it more susceptible to low reflectivities \cite{Machado}. Furthermore, while low-gain PDC is typically driven by continuous-wave lasers, high-gain PDC relies on pulsed lasers, potentially enabling time-gated measurements. \par
\begin{figure}[!ht]
\includegraphics[scale=0.35]{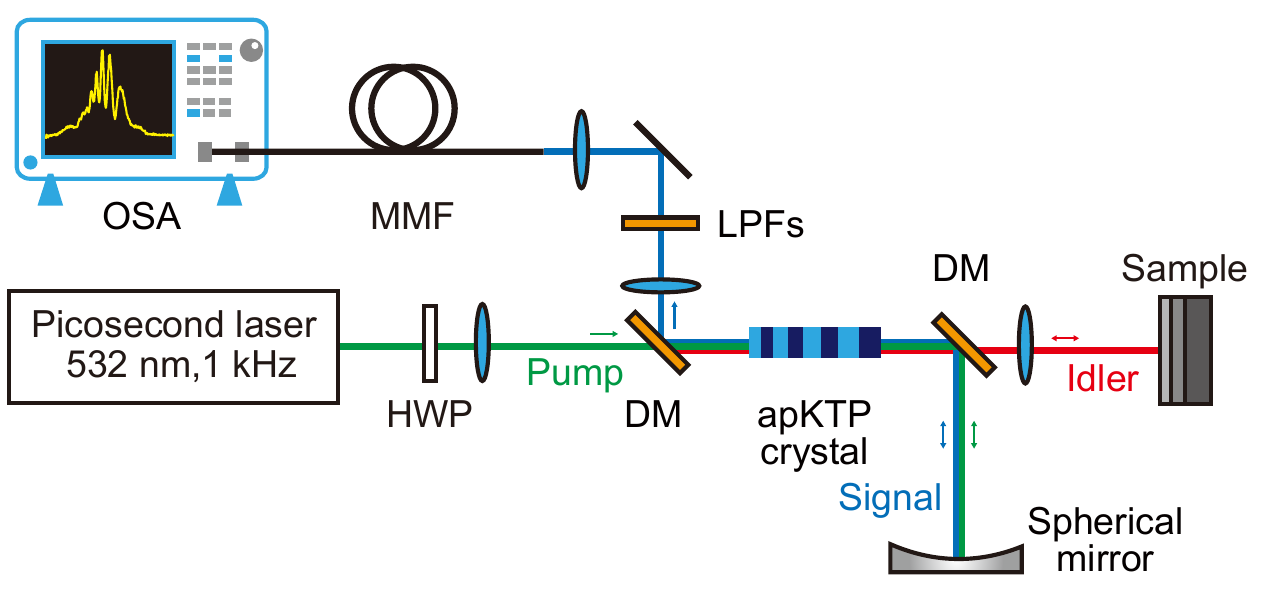}
\centering
\caption{\textit{Schematic of LCI with undetected {MIR} photons. HWP: Half-wave plate; DM: Dichroic mirror; LPF: Long-pass filter; MMF: Multimode fiber; OSA: Optical spectrum analyzer}}
\label{fig:setup}
\end{figure}

Previous implementations of LCI with undetected photons in the high-gain regime have been limited to the visible and NIR ranges \cite{Horoshko2020,Machado,Hashimotohighgainvis}. 
In this work, we demonstrate for the first time LCI with undetected MIR photons in the high-gain regime. We employ broadband, non-degenerate PDC in an apKTP crystal to generate a MIR idler beam centered at $3$ $\mu$m and a visible signal beam centered at circa $600$ nm. Two crystal designs are considered, differing in the the range of variation of the poling period and thus the PDC bandwidth, yielding axial resolutions of $30$ $\mu$m and $17$ $\mu$m, respectively. The experimental setup is schematically illustrated in Fig. \ref{fig:setup}.

The pump source for our SU(1,1) interferometer is a 15-ps pulsed laser, centered at $532$ nm and with a repetition rate of $1$ kHz. The pump beam is focused on a type-0 apKTP crystal using an
$f = 200$ mm lens to induce collinear high-gain PDC. The crystal poling frequency changes along the crystal itself, a technique known to increase the bandwidth compared to standard periodic poling \cite{Chekhovaapktp,HoroshkoapKTP}, allowing to access sufficiently broadband PDC even for highly non-degenerate frequencies. Two different crystal designs are used, differing in their poling period ranges ($\Lambda_1$ = 12.3-14.0 $\mu$m, $\Lambda_2=$10.9-14.0 $\mu$m). They are used to generate a broadband MIR idler (centered at $3.3$ and $2.6$ $\mu$m, respectively) and a visible signal (centered at $635$ and $670$ nm, respectively). After the crystal, a dichroic mirror separates the generated signal and the residual pump from the idler beam. The idler pulse reaches the LCI sample after being collimated by an $f = 100$ mm $\mathrm{CaF_2}$ lens. Its waist at the sample position is measured to be $(3.3\pm0.5)$ mm, as detailed in Supplementary Fig. S3. The back-reflected idler is then redirected into the nonlinear crystal, along with the pump and signal beams, which are reflected by a spherical mirror. 
A motorized delay stage in the sample arm enables fine-tuning of the optical path length difference between the two arms. During the second pass, the signal pulse is amplified/deamplified through the OPA process, depending
on the pump phase relative to the signal and idler phases \cite{Hashimotohighgainvis, Hashimoto2024}. The resulting interference, encoded in the modulated signal spectrum, contains information about the optical path length traveled by the idler beam, and thus about the sample's depth structure. After the crystal, the signal is isolated from the other pulses using a second dichroic mirror followed by long-pass filters. It is then coupled to a multimode fiber with a $200$ $\mu$m core diameter and analyzed using an optical spectrum analyzer (OSA). For a single reflective layer in the sample, the measured signal spectrum can be modeled as \cite{Hashimotohighgainvis}:
{
\begin{equation}
S(\omega_s)=S_0(\omega_s)\left[1+\mathcal{V}_0\cos\left(2z\frac{\omega_p-\omega_s}{c}+\rho(\omega_s)\right)\right],
\end{equation}
}
where $\omega_s$ is the signal angular frequency, $S_0(\omega_s)$ the non-interference signal spectrum, { $\mathcal{V}_0$ the visibility, $z$ the axial position of the sample, $\omega_p$ the pump angular frequency, {$c$ the speed of light in vacuum}, and $\rho(\omega_s)$ the chirped phase determined by the group velocity dispersion and the poling profile of the crystal \cite{Hashimotohighgainvis}.} To recover the depth profile of the sample, we apply a Fourier transform to the measured signal spectrum {and map the delay time $t$ to the sample depth $z'$ by $t=2z'/c$}, thus achieving Fourier-domain (FD) LCI. The full data processing procedure is described in { Supplement} {1}.

\begin{figure}[!ht]
\includegraphics[scale=0.35]{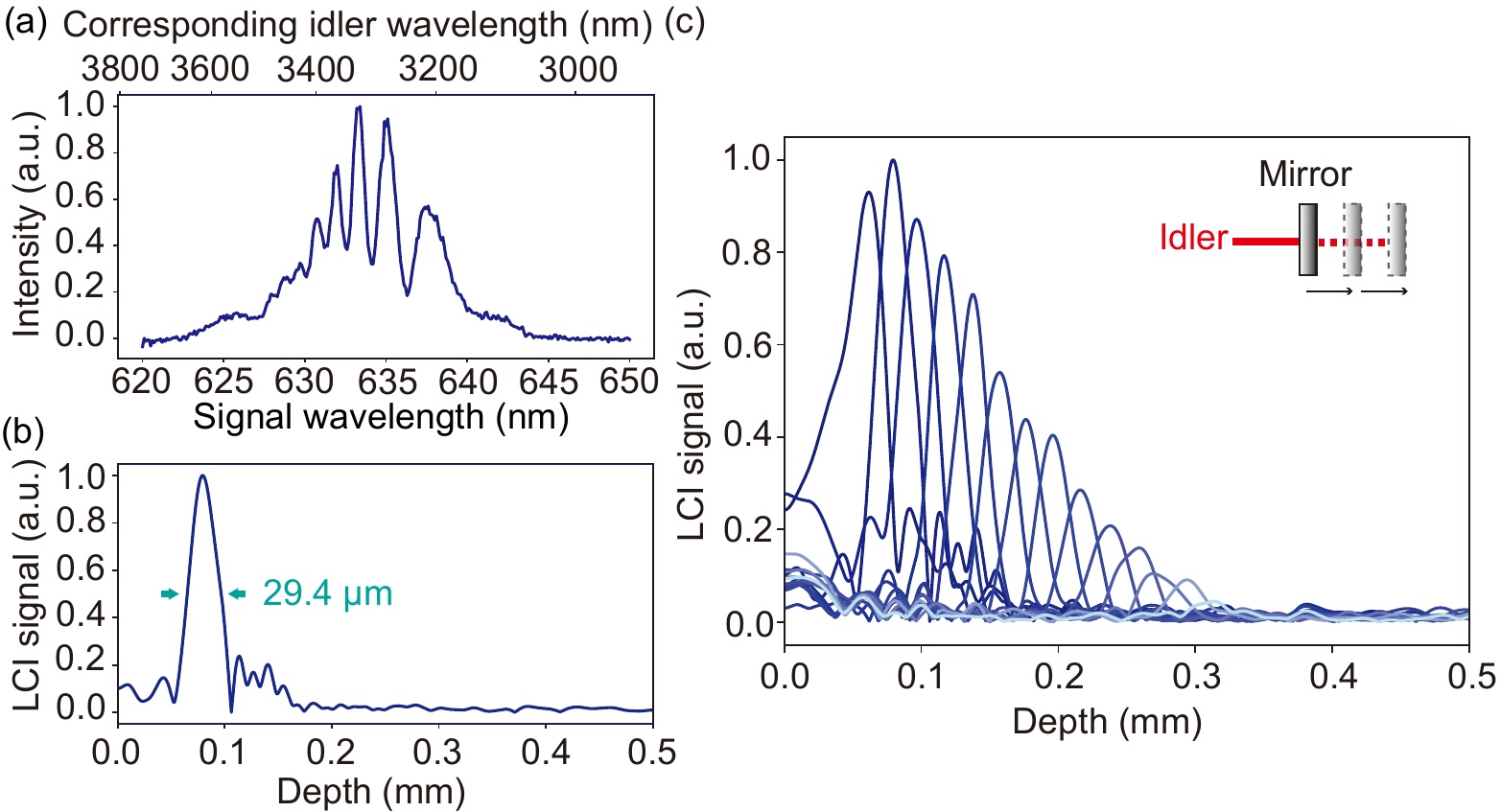}
\centering
\caption{\textit{(a): Signal spectrum at the output of the interferometer, for a fixed position of the mirror along the sample arm. The interference pattern is related to the optical path difference between pump, signal, and idler beams; (b): Resulting depth profile, with the position of the peak corresponding to the one of the sample; (c): Reconstructed roll-off curve, measured by scanning several axial positions of the mirror.}}
\label{fig:characterization}
\end{figure}

We characterize the interferometer performance using the first apKTP crystal design. The results are reported in Fig. \ref{fig:characterization}.

Fig. \ref{fig:characterization}(a) shows the output signal spectrum acquired using a silver mirror as the sample, exhibiting a full width at half maximum (FWHM) bandwidth of approximately $10$ nm. Fig. \ref{fig:characterization}(b) displays the resulting LCI depth profile, from which we can infer an axial resolution equal to $(29.4\pm0.2)$ $\mu$m, comparable with the simulation results (see Supplementary Fig. S4). The pump power is set to $1$ mW, corresponding to a detected signal power equal to $(9.0\pm0.5)$ nW. The signal-to-noise ratio (SNR) is also measured according to the standard definition \cite{drexler2008optical}
\begin{equation}        
\mathrm{SNR}=20\log\left({\frac{I_\text{LCI}}{\delta I_\text{LCI}}}\right),
\end{equation}
with {$I_\text{LCI}$} being the LCI signal peak, {$\delta I_\text{LCI}$} the standard deviation of the background. An SNR of approximately $40$ dB is found. At this power level, the dominant noise source is identified as relative intensity noise (RIN), rendering the SNR independent of the detected signal power (see Supplementary Fig. {S5}). Fig. \ref{fig:characterization}(c) shows the measured roll-off curve, reconstructed by scanning several axial positions of the mirror. The step-by-step distance is set to $20$ $\mu$m, in good agreement with the one exhibited by the corresponding LCI peaks. The LCI signal is found to drop by $10$ dB at a distance around $270$ $\mu$m from the zero-path difference position. The decrease in the level of SNR with the sample distance is inevitable in FD LCI, and it is associated with the ability to resolve channeled spectrum oscillations \cite{rolloff}, becoming finer and finer with distance. When using continuous-wave lasers as the source, their coherence length is sufficiently long that the depth range is primarily determined by the detector's spectral resolution. In contrast, the finite spectral width of pulsed lasers can lead to imperfect frequency correlations, possibly affecting the depth range of the FD LCI measurement~\cite{Takeuchi_rolloffpulsed}. In our case, simulations show that accounting for the finite spectral resolution of the OSA, $0.6$ nm, is enough to explain the experimental roll-off (see Supplementary Fig. {S4}). We can then conclude that the pump coherence length is long enough not to affect the resultant depth range. 

The axial resolution can also be affected by the roll-off. Fig. \ref{fig:characterization}(c) shows that the FWHM of the LCI peaks is quite constant between $0.1$ and $0.3$ mm from the zero-delay position, but somewhat broadens at shorter and longer distances. Most likely, the worsening of the axial resolution below $0.1$ mm is related to the presence of a higher residual DC component, affecting the quality of the Fourier transform. \par
\begin{figure}[!ht]
\includegraphics[scale=0.35]{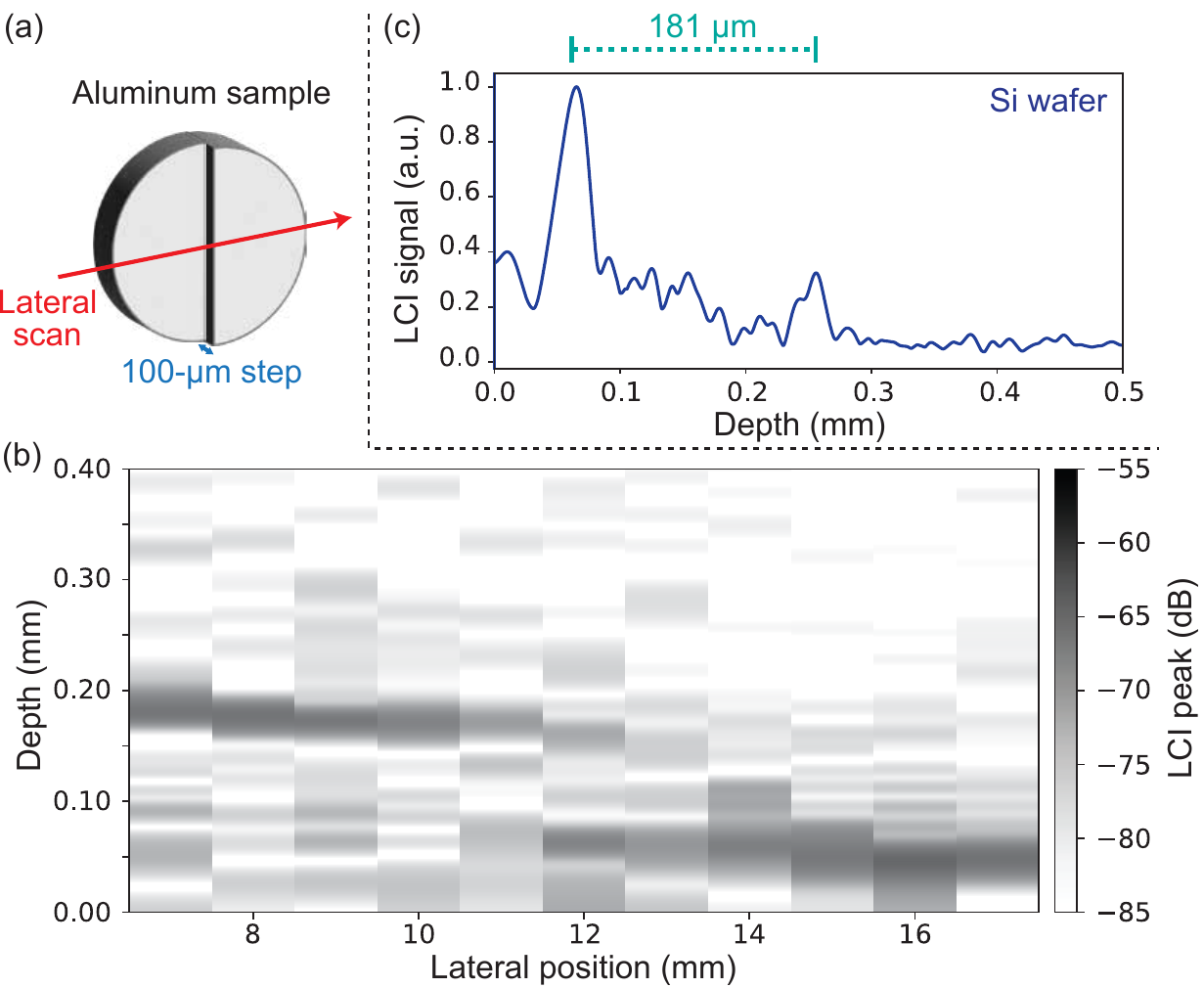}
\centering
\caption{\textit{(a): 3D schematic of Al-stepped sample; (b): lateral scan of the Al-stepped sample; (c): Reconstructed depth profile of the thin Si layer, placed behind a Ge window.}}
\label{fig:samples}
\end{figure}

To demonstrate the imaging capabilities of the setup, we first reconstruct the 2D profile of an aluminum step mirror, with a nominal step height of $100$ $\mu$m. Fig. \ref{fig:samples} (a) shows the sample schematic, while Fig. \ref{fig:samples} (b) displays its measured 2D map, showing the depth profiles corresponding to different lateral positions. The darkest thick lines (with the thickness given by the axial resolution) correspond to the two halves of the reflective surface, indeed at different depths because of the step structure. The step height can be estimated by measuring the distance between such lines, carefully accounting for their tilting due to the sample being non-perfectly normal to the beam. A value of $(100 \pm 3)$ $\mu$m is found.\par

We further prove the MIR LCI capability of the setup by measuring a second sample, a Si layer of nominal thickness $50$ $\mu$m, after placing a $1\,\mathrm{mm}$ thick BBAR-coated Ge window in front of it, along the sample arm. The window blocks nearly all light below $2$ $\mu$m, making standard visible/NIR-OCT imaging infeasible. Such Ge layers (thickness 150–200 $\mu$m) are common in electronics applications, such as Ge-based solar cells \cite{solarcellthick}. Fig. \ref{fig:samples}(c) shows the Si-layer LCI scan, with the two peaks corresponding to the front and back surfaces, respectively. The optical distance between them, $(181 \pm 3)$ $\mu$m, corresponds to the thickness of $(52.2 \pm 0.9)$
$\mu$m accounting for the refractive index of Si, consistent with the manufacturer's specification ($50\pm10$) $\mu$m.\par
To prove the two-pass signal amplification advantage enabled by the high-gain regime, we compare the signal power at the output of the interferometer with the estimated idler power sent into the sample. The signal and idler beams generated in the PDC process exhibit photon-number correlations \cite{photonnumbercorr}. This means that for balanced losses, we expect the idler and signal to have the same number of photons at conjugated wavelengths. Therefore, we can estimate the idler spectrum after the first pass simply by mirroring the signal spectrum around the frequency $\omega_p/2$. Finally, we can convert the photon spectrum into power. To measure the signal spectrum after the first OPA, we deviate the signal beam from its usual path in Fig. \ref{fig:setup} with the help of a flip  mirror. We collimate and then focus it on an optical fiber, directly connected to a visible spectrometer (with a spectral resolution of $1.4$ $\mathrm{nm}$). Meanwhile, the signal after the second OPA is measured with the same spectrometer after traveling the usual path. Comparison of the spectra from the two stages (Fig. \ref{fig:amplif}) shows that they differ by more than an order of magnitude.

\begin{figure}[!ht]
\includegraphics[scale=0.4]{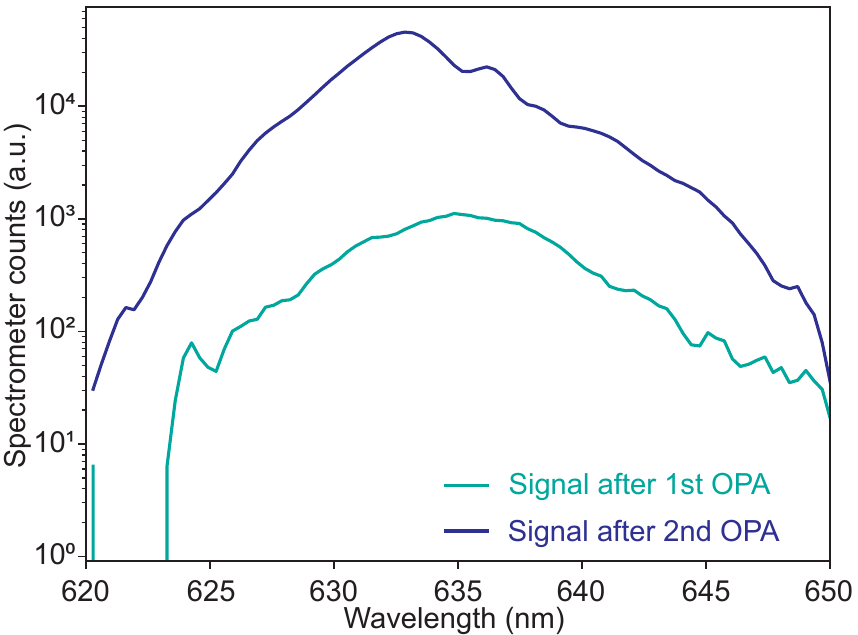}
\centering
\caption{\textit{Comparison between the signal spectrum after the first (blue-green) and the second (blue) passage through the nonlinear crystal, in log scale.}}
\label{fig:amplif}
\end{figure} 

Considering a band of signal frequencies around $\omega_s$ and the matching band of idler frequencies around $\omega_i=\omega_p-\omega_s$, we can approximate the ratio between the detected power (signal after the second pass) and the probing power (idler after the first pass) as \cite{Hashimotohighgainvis}
\begin{equation}
{\frac{P_\text{signal,2}}{P_\text{idler,1}}}\approx\frac{1}{2}\frac{\sinh^2(2r)}{\sinh^2(r)}\frac{\omega_s}{\omega_i},    
\end{equation}
where $r$ is the parametric gain, assumed to be the same at both stages. In the low-gain regime, $r\ll1$, therefore ${\frac{P_\text{signal,2}}{P_\text{idler,1}}}\approx2\frac{\omega_s}{\omega_i}$. In contrast, in the high-gain regime $r\gg1$, thus ${\frac{P_\text{signal,2}}{P_\text{idler,1}}}\approx\frac{1}{2}e^{2r}\frac{\omega_s}{\omega_i}$, with an enhancement factor $e^{2r}/4$ over the low-gain regime. Experimentally, we find the amplification to be as high as $208$, an order of magnitude larger than what would be achieved in the low-gain regime under the same conditions ($\approx 10.3$) and what has been indeed reported in the literature for the low-gain regime ($\approx 9.6$ in \cite{Vanselow:20}). 
This measurement is performed for different pump powers above $1.5$ $\mathrm{mW}$, the minimum required to detect the first-pass signal spectrum. The complete dependence of the amplification factor is reported in Supplementary Fig. {S6}.\par
\begin{figure}[!ht]
\includegraphics[scale=0.35]{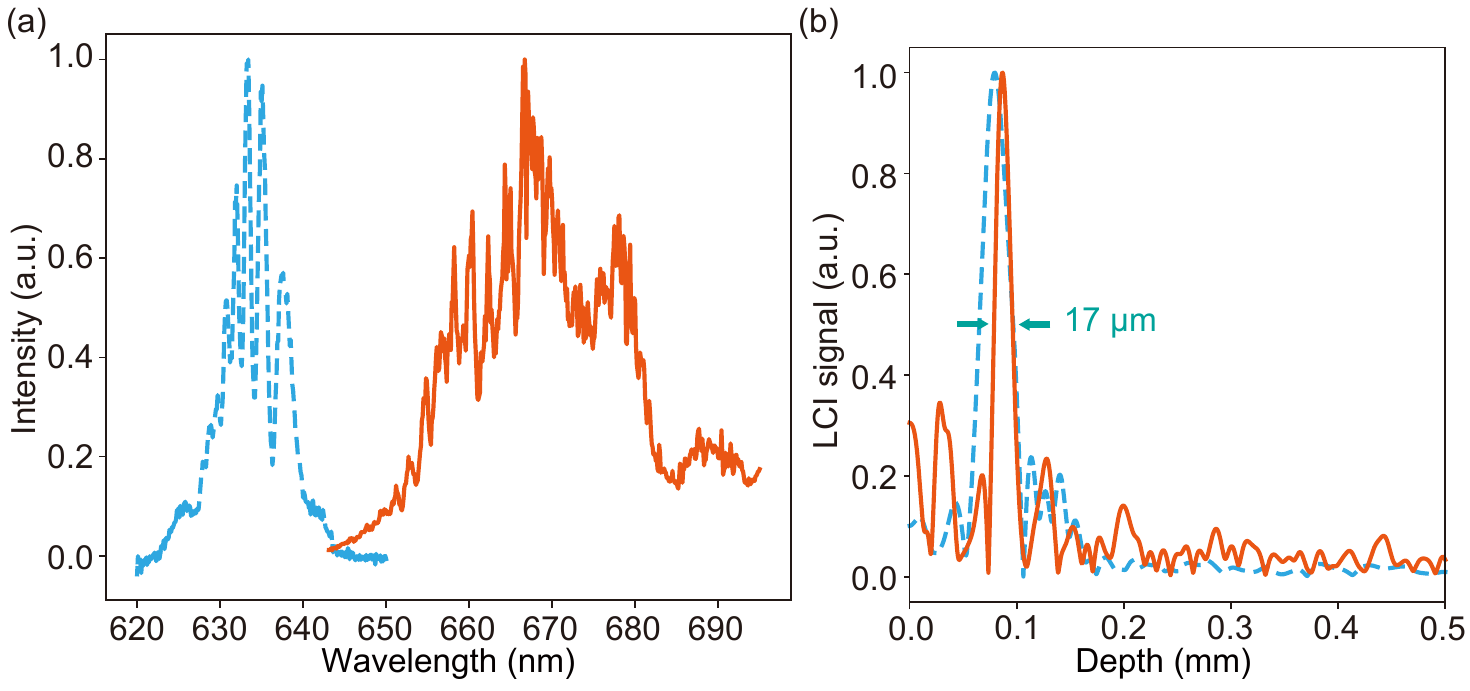}
\centering
\caption{\textit{(a): Comparison between the output signal spectra for the crystal design with $\Lambda_1=12.3-14.0$ $\mu$m (dotted sky-blue line) and the one with $\Lambda_2=10.9-14.0$ $\mu$m (solid orange line), obtained for similar positions of the mirror; (b): Comparison between the corresponding LCI spectra, showing the improvement in the axial resolution.}}
\label{fig:sktp}
\end{figure}

Finally, we repeat the performance characterization for the second crystal design, which is engineered to generate a broader PDC spectrum and thus provide a better axial resolution. The comparison between the two interference spectra and the corresponding LCI peaks is reported in Fig. \ref{fig:sktp}(a) and Fig. \ref{fig:sktp}(b), respectively. The dotted sky-blue line shows the results obtained for the first crystal design, while the solid orange line shows the results for the second design.

For similar mirror positions, we observe a doubling of the signal bandwidth, from $10$ nm to $22$ nm, resulting in an enhanced axial resolution of $17$ $\mu$m. This improvement comes at the cost of reduced SNR, which decreases to $32$ dB at a pump power of $(7.6 \pm 0.5)$ mW. These results confirm that the axial resolution can be improved by increasing the range of the poling period.

In summary, we report the first MIR LCI setup based on high-gain SU(1,1) interferometry with undetected photons. We exploit the PDC process inside an apKTP crystal, generating a broadband {MIR} idler, centered around $3$ $\mu$m, and a visible signal, centered around $0.6$ $\mu$m. The system provides solid MIR LCI performance, achieving SNR as high as $40$ dB, axial resolution around $30$ $\mu$m, and depth range around $270$ $\mu$m. These results come together with the advantages of operating in the high-gain regime, both in terms of the amount of detected power, around $9$ nW, and {the amplification of the photon flux between probing the object and detecting the signal.} The ratio between the probing power and the detected power is indeed found to be higher than $200$, {one order} of magnitude larger than what is achieved in the low-gain regime. Furthermore, we improve the axial resolution of our system to $17$ $\mu$m, corresponding to a factor 2 enhancement, by modifying the crystal design, extending the range of the poling period. We thus demonstrate a practical route to enhance the axial resolution via the engineering of the crystal poling profile.\par
Future applications of our work include non-invasive MIR imaging of highly scattering materials, i.e. ceramic coatings and oil paintings, as well as of devices containing semiconductors with transmission window in the infrared. The use of a pulsed source also paves the way for time-gated measurements,  potentially enabling dynamic sample studies. Looking ahead, a possible improvement of the setup involves focusing the idler beam on the sample, to achieve OCT. Another promising direction is further engineering of the nonlinear crystal, e.g., using apodized and chirped poling profiles to optimize bandwidth and spectral flatness, which could enhance axial resolution without sacrificing visibility. 

\begin{backmatter}
\bmsection{Funding} 
Collegio Superiore Bologna (Italy) (DM231/2023), JST PRESTO (Japan) (JPMJPR2457), Deutsche Forschungsgemeinschaft (CH 1591/16-1), Bundesministerium für Bildung und Forschung (Germany) (13N16935), Horizon 2020 Framework Programme (European Union) (101017733).
\bmsection{Acknowledgements} 
The authors thank Kyoohyun Kim for letting us use the microscope to inspect crystals, Robert Gall for preparing Al-stepped samples, and Irina Harder for supporting Si sample preparation.
\bmsection{Disclosures} The authors declare no conflict of interest.
\bmsection{Data availability} The data provided in the manuscript are available from the corresponding author upon reasonable request.

\end{backmatter}

\title{Supplementary information}
\section{Theoretical model}

Several samples of aperiodically poled potassium titanyl phosphate (apKTP) were manufactured (by Raicol Crystals), {and part of them were} used in the experiments: a short-range crystal (local period $\Lambda_1$ = 12.3-14.0 $\mu$m) with a linear poling profile and a long-range crystal (local period $\Lambda_2=$10.9-14.0 $\mu$m) with a logarithmic profile. 

\begin{figure}[!ht]
\includegraphics[width=0.8\textwidth]{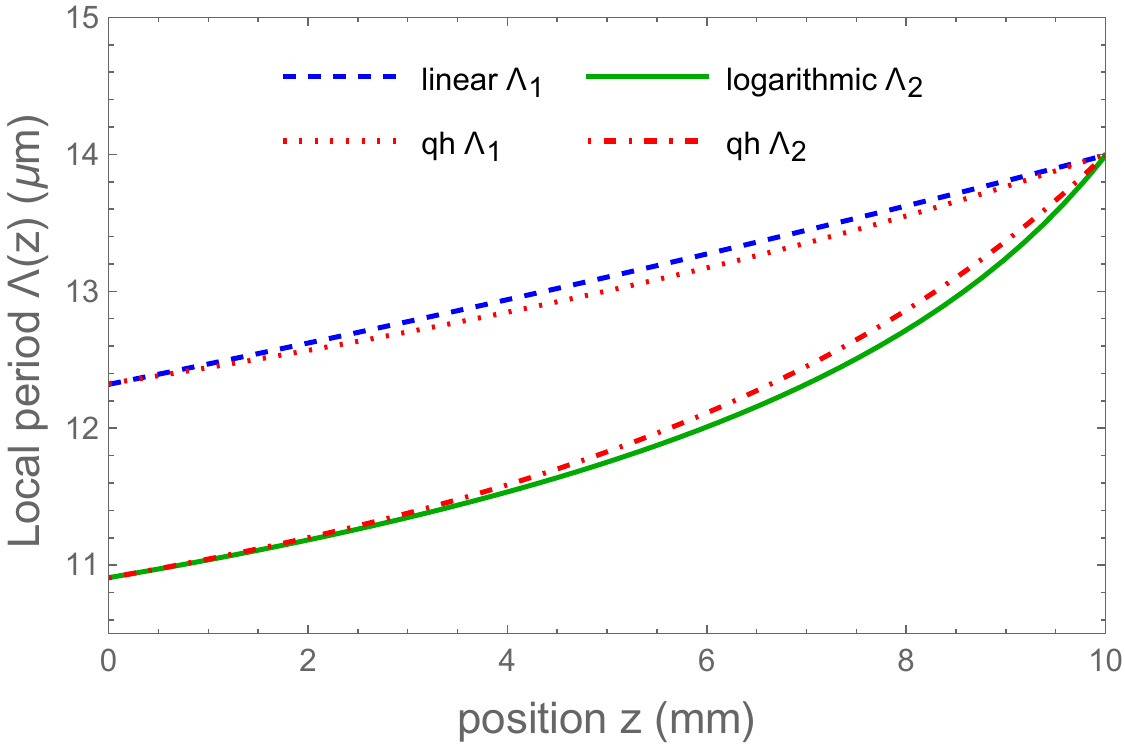}
\centering
\caption{\textit{{ Several poling profiles of apKTP designed for this work. The quadratic-hyperbolic (qh) profiles (dotted and dash-dotted red) are close to the linear (dashed blue) or logarithmic (solid green) profiles with the same local period range. The two crystals with the linear and logarithmic profiles are used in the experiments}}}
\label{fig:profiles}
\end{figure} 

The idea behind using a nonlinear poling profile is to avoid spectral narrowing caused by nonlinear loss in the crystal: The crystals are designed to have a more gentle slope of the function $\Lambda(z)$ in the beginning and a more steep slope at the end. Since the intensity of the generated light is inversely proportional to the derivative $\partial\Lambda/\partial z$, the intensities of the signal and idler fields generated in the beginning of the crystal are higher than those generated in the end. This asymmetry is intended to counterbalance the asymmetry introduced by the nonlinear loss \cite{Hashimoto2024}. 

The theoretical model will be developed for the quadratic-hyperbolic profile, well studied in Refs. \cite{HoroshkoapKTP,Horoshko2020,Hashimoto2024}. This profile has the advantage of characteristic phases in the form of second-order polynomials of the frequency, which simplifies the analysis. As we can see in Fig.~\ref{fig:profiles}, the difference between the quadratic-hyperbolic profiles and the reported profiles that have the same range is not large, and the results of our analysis will be applicable to the results reported in the main text.  

The first crystal (local period $\Lambda_1$ = 12.3-14.0 $\mu$m) is quasi-phase-matched for the spectral range between $\lambda_\text{high}=664$ nm and $\lambda_\text{low}=620$ nm. However, only a part of this range between $\lambda_\text{up}=640$ nm and $\lambda_\text{down}=630$ nm was observed. We ascribe this spectral narrowing to nonlinear loss due to simultaneously quasi-phase-matched nonlinear processes, mainly sum-frequency generation from the pump and the signal or idler fields \cite{Hashimoto2024}. {A higher steepness gradient may mitigate the narrowing effect.} We model the resulting spectrum $S_0(\omega_s)$ by a Gaussian distribution of circular frequency centered at
{ $\omega_{s0}=\frac12(\omega_\text{down} + \omega_\text{up})$ and having the standard deviation of $\Sigma=(\omega_\text{up} - \omega_\text{down})/2\sqrt{2\ln2}=2\pi\times3.2$ THz, where $\omega_\text{down}=2\pi c/\lambda_\text{up}$, $\omega_\text{up}=2\pi c/\lambda_\text{down}$, and $c$ is the speed of light in vacuum.
}

The signal spectrum at the interferometer output is given by Eq. (1) of the main text. The phase $\rho(\omega_s)$ at signal circular frequency $\omega_s$ in this equation is determined by the poling profile and the group velocity dispersion of the crystal and has the form \cite{Horoshko2020,Hashimotohighgainvis} 
\begin{equation}\label{rho}
\rho(\omega_s) = -\frac{2\alpha L}{\omega_0^2}\frac{\omega_\text{low}-\omega_0}{\omega_\text{high}-\omega_\text{low}}(\omega_s - \omega_\text{high})^2,
\end{equation}
where $\omega_p$ is the pump circular frequency, $\omega_0=\omega_p/2$, $\omega_\text{low}=2\pi c/\lambda_\text{high}$, $\omega_\text{high}=2\pi c/\lambda_\text{low}$, and $L$ is the crystal length. { Here we assume that, within the signal band, the approximation holds $\Delta k(\omega_s)\approx-\frac1{\omega_0^2}\alpha (\omega_s-\omega_0)^2+\beta$ with $\alpha=406$ rad/mm and $\beta=657$ rad/mm, where $\Delta k(\omega_s)=k(\omega_p)-k(\omega_s)-k(\omega_p-\omega_s)$ is the phase mismatch of the parametric down-conversion, determined by the dispersion law $k(\omega)$ of the $Z$-polarized wave propagating along the $X$-axis of the KTP crystal.} Equation (\ref{rho}) represents a parabola with its vertex at $\omega_\text{high}$ that is quasi-phase-matched on the crystal output face. We rewrite this equation as
\begin{equation}\label{rho2}
\rho(\omega_s) = -2z_0 \frac{\omega_p-\omega_s}c - \frac{(\omega_s - \omega_{s0})^2}{2\omega_\text{ch}^2} +\rho_0,
\end{equation}
where 
\begin{equation}\label{chirp}
\omega_\text{ch} = \frac{\omega_0}2 \sqrt{\frac{\omega_\text{high} - \omega_\text{low}}{\alpha L(\omega_\text{low}-\omega_0)}} = 2\pi\times0.96\,\text{THz}
\end{equation}
is the chirp scale, $\rho_0$ is an irrelevant constant phase, and
\begin{equation}\label{z0}
z_0 = c\frac{\omega_\text{high} - \omega_{s0}}{2\omega_\text{ch}^2} = 295\,\mu\text{m}
\end{equation}
is the sample depth at which the signal photon at frequency $\omega_{s0}$ arrives at its amplification layer (in the second pass) simultaneously with its twin idler photon at frequency $\omega_p-\omega_{s0}$. 
{ Alternatively, $z_0$ can be expressed as 
\begin{equation}\label{z0bis}
z_0 = c\left(\frac{L-z_\text{pm}(\omega_{s0})}{v_g(\omega_{s0})} - \frac{L-z_\text{pm}(\omega_{s0})}{v_g(\omega_p-\omega_{s0})} \right),
\end{equation}
where $z_\text{pm}(\omega)$ is the perfect phase-matching position for frequencies $\omega$ and $\omega_p-\omega$ and $v_g(\omega)$ is the group velocity at frequency $\omega$, see Eq. (61) of Ref. \cite{HoroshkoapKTP}.

Note that a crystal with a linear poling profile has additional terms of orders 3 and 4 in $\omega_s$ in the expression for $\rho(\omega_s)$, Eqs. (\ref{rho}) and (\ref{rho2}), even if the dispersion law $k(\omega)$ is assumed to be quadratic in frequency, see Eq. (67) of Ref. \cite{HoroshkoapKTP}. However, these additional terms are very small within the observed spectral bandwidth and can be neglected. Similarly, a crystal with a logarithmic poling profile has additional terms of all orders in the expression for $\rho(\omega_s)$, but the terms of orders greater than 2 can be disregarded within the observed spectral bandwidth. These small variations of $\rho(\omega_s)$ are the only difference between the quadratic-hyperbolic profiles of our analytical model and the profiles used in the experiment.
}

\begin{figure}[!ht]
\includegraphics[width=0.49\columnwidth]{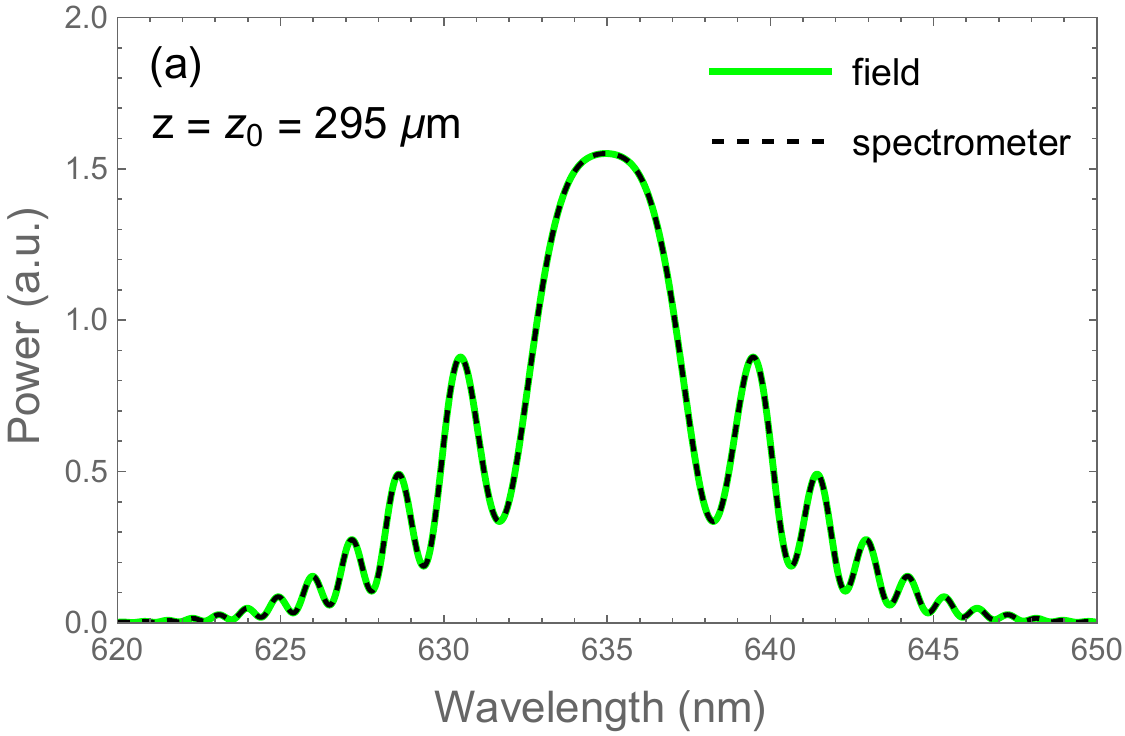}
\includegraphics[width=0.49\columnwidth]{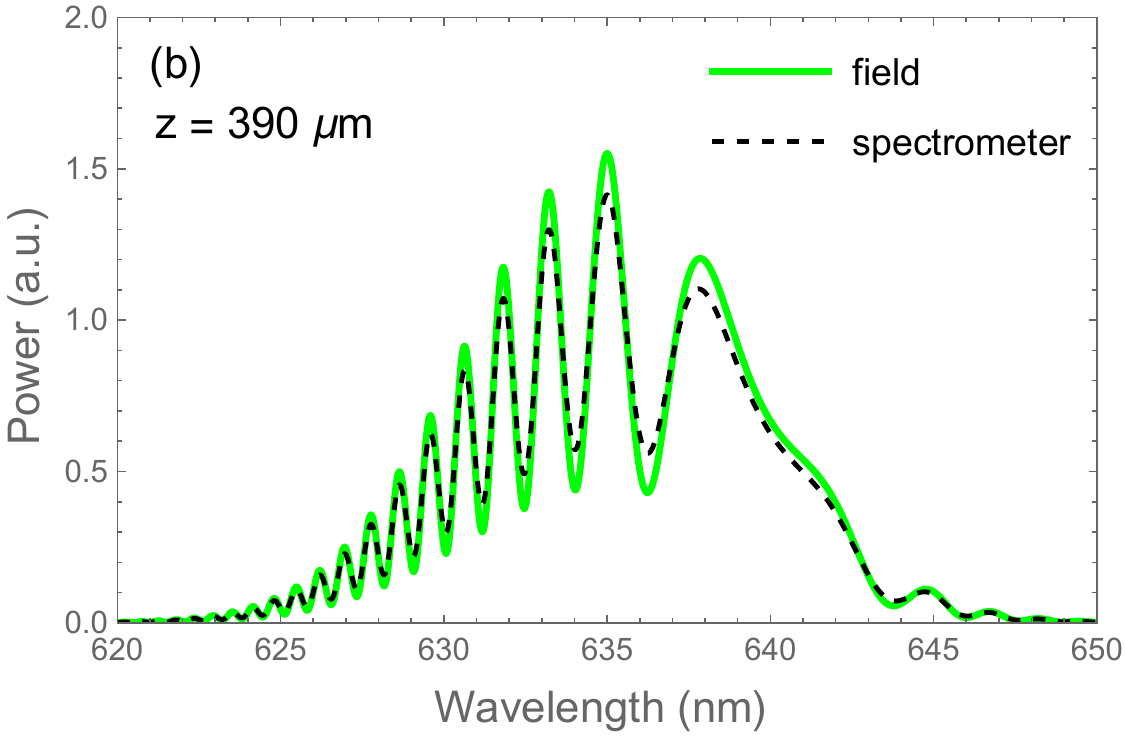}
\centering
\caption{\textit{Modeled signal spectra at the interferometer output for visibility ${\mathcal{V}_0}=0.55$ and two different sample depths $z$ {((a): $295$ $\mu m$, (b): $390$ $\mu m$)}. The spectrum of the field {(solid light green)} is given by Eq. (1) of the main text, while the spectrum of the spectrometer {(dashed black)} is given by Eq. (\ref{Sobs}).}}
\label{fig:spectra}
\end{figure}

The observed spectrum is the result of a convolution of the field spectrum with the instrumental function of the spectrometer, which we model by a Gaussian distribution of circular frequency with a standard deviation of $\sigma=\delta\lambda\omega_{s0}^2/ 4\pi c \sqrt{2\ln2} =2\pi\times0.19$ THz, where $\delta\lambda=0.6$ nm is the resolution of the spectrometer. In the limit $\sigma\ll\Sigma,\omega_\text{ch}$, we obtain
\begin{equation}\label{Sobs}
S_\text{spec}(\omega_s) = S_0(\omega_s)\left(1+\mathcal{V}(z)\cos\left[2(z-z_0)\frac{\omega_p-\omega_s}{c} - \frac{(\omega_s - \omega_{s0})^2}{2\omega_\text{ch}^2} +\rho_0\right]\right),
\end{equation}
where $\mathcal{V}(z) = \mathcal{V}_0 \exp\left[-(z-z_0)^2/2\sigma_V^2\right]$ is the depth-dependent visibility with its standard deviation $\sigma_V=c/2\sigma=126$ $\mu$m. The modeled spectra for two values of the sample depth $z$ are shown in Fig.~\ref{fig:spectra}. The second modeled spectrum resembles the measured one reported in Fig.~2(a) of the main text. 

The spectrum in Eq.~(\ref{Sobs}) is a sum of two terms, where the first is independent of the sample depth. Measuring the spectrum at depth $z\gg\sigma_V$, where $\mathcal{V}(z)\approx0$, we obtain the spectral background $S_0(\omega_s)$. Subtracting this background from the measured spectrum and taking the inverse Fourier transform $\omega_s\to2z'/c$, we obtain a sum of two peaks centered at $z'=\pm(z-z_0)$. Taking the modulus of the peak with the $+$ sign, we obtain the low-coherence interferometry (LCI) signal 
\begin{equation}\label{Ilci}
I_\text{LCI}(z'|z) = I_0\mathcal{V}(z) \exp\left[-(z'-z+z_0)^2 /2\sigma_z^2\right],
\end{equation}
where $I_0$ is a constant independent of $z$ and $z'$ and $\sigma_z$ is the standard deviation given by 
\begin{equation}\label{sigmaz}
\sigma_z = \frac{c}{2\Sigma}\sqrt{1+\frac{\Sigma^4}{\omega_\text{ch}^4}} = 82\,\mu\text{m}.
\end{equation}
The standard deviation of the LCI signal is the product of two factors, the inverse spectral width $c/2\Sigma=7.6$ $\mu$m, and the chirping broadening factor $\sqrt{1+(\Sigma/\omega_\text{ch})^4}\approx11$. The second factor can be removed by proper signal processing equivalent to putting $\omega_\text{ch}\to\infty$, giving the theoretical FWHM width of the LCI signal peak of $\Delta z=\sqrt{2\ln2}c/\Sigma=18$ $\mu$m. {Alternatively, the spectral phase associated with the second factor can be removed by placing a {slab of} glass in the sample arm.}

\section{Data analysis procedure}

The data analysis procedure adopted to reconstruct the LCI spectra from the raw interference patterns and consequently evaluate the {SNR} is the following:
\begin{enumerate}
    \item we preliminarily evaluate the non-interference background for each sample, by averaging a total of 20 measurements;
    \item we divide each interference spectrum by the corresponding background, to isolate the interference pattern and remove the envelope contribution. We also subtract the average value from the result, in order to remove the constant part responsible for the DC component in the Fourier transform;  
   
    \item we restrict the operative range to the region where the interference fringes are clearly visible, to minimize high-frequency and low-frequency noise. This region is found to be $623-645$ $\mathrm{nm}$ for the first crystal design, $641.5-680$ $\mathrm{nm}$ for the second one;
    \item we convert data from the wavelength domain to the frequency domain, via interpolation to preserve equal spacing;
    \item at this stage, processed data might still suffer from nonlinear terms in the phase, resulting from group-velocity dispersion between the beams (accounted for in the {chirped phase $\rho(\omega_s)$ in Eq. (1) in the main text)} \cite{Vanselow:20}. To compensate for this, we apply the Hilbert transform to the data and multiply by a phase term of the form {$e^{-i\eta (\omega_{s}-\omega_{s0})^2 }$} \cite{Hilbert}, where 
    $\eta$ is a dispersion coefficient to be determined experimentally;
    \item we apply both windowing and zero-padding to the Hilbert-transformed spectra. In particular, the standard Hanning window function is chosen for apodization of the interference patterns \cite{proHanning}. The Hanning window provides optimal performance in terms of achievable SNR, but it is also associated with a significant broadening of the achievable axial resolution \cite{Hanninglimit}. This explains the discrepancy between the estimated achievable axial resolution values derived in Supplement 1 and the ones obtained experimentally as well as in the simulations.
    
    \item we finally apply the inverse Fast Fourier Transform and convert the result from the time domain to the depth domain.
\end{enumerate}

\section{Lateral resolution measurement}
\begin{figure}[!ht]
\includegraphics[scale=0.5]{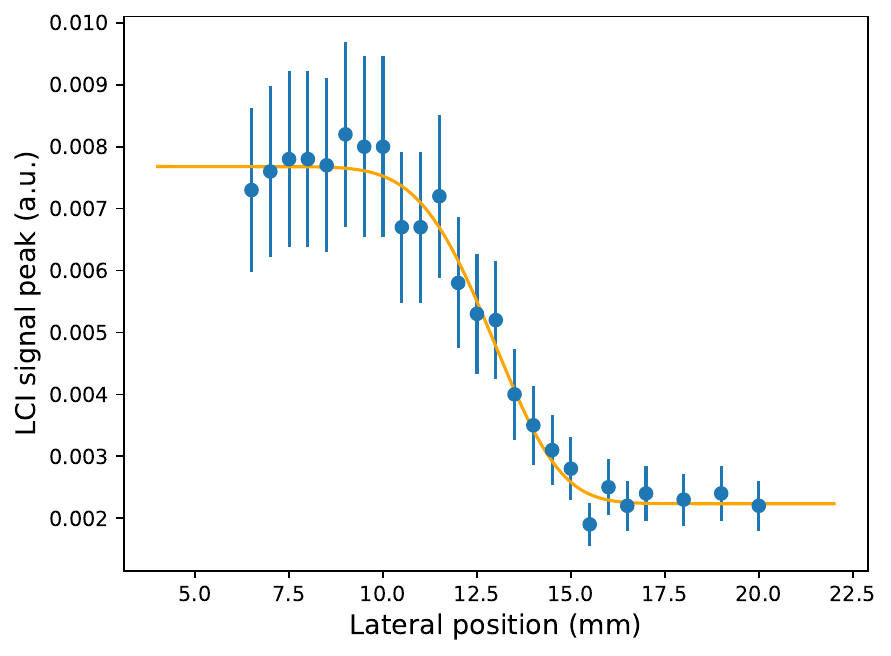}
\centering
\caption{\textit{The dependence of the LCI signal on the lateral position for the D-
shaped mirror. The orange line corresponds to the fit with the expected $\mathrm{\sqrt{erfc}}$ behavior. The error bars correspond to the statistical uncertainties.}}
\label{fig:knifeedge}
\end{figure}

Since the idler beam is collimated and sent into the sample with a macroscopic size, the lateral resolution is limited by its diameter. To estimate it, we perform a knife-edge-like measurement, using a D-shaped mirror as the sample. While scanning the mirror laterally at a fixed depth, we expect the total reflected power to change as an $\mathrm{erfc}$ function of the lateral position. The beam's diameter can be estimated as the FWHM of the derivative of such a function. The LCI signal dependence on the lateral resolution can be also used to reconstruct this erfc behavior. In general, the LCI signal depends on the output power {$P_\text{out}$} of the interferometer as \cite{kalkman}
\begin{equation}
    I_{LCI}\propto {P_\text{out}\mathcal{V}_0}
    \label{eq:lci_sig}
\end{equation}
The visibility in the high-gain regime scales nonlinearly with the sample's reflectivity, namely \cite{Hashimotohighgainvis}
\begin{equation}
    {\mathcal{V}_0}=\frac{2\sqrt{R_rR_s}}{R_r+R_s},
\end{equation}
where $R_r$ and $R_s$ are the reflectivities of the reference and sample arms, respectively. A shift in the lateral position of the D-shaped mirror is mathematically analogous to a change in the sample's reflectivity. Therefore, we can rewrite Eq. \ref{eq:lci_sig} as a function of the lateral position $d$
\begin{equation}
    {I_\text{LCI}(d)\propto P_\text{out}(d)\mathcal{V}_0(d)}=(R_s(d)+R_r)P_0\frac{2\sqrt{R_s(d)R_r}}{R_s(d)+R_r},
\end{equation}
where $P_0$ is the output power when all the idler beam is backreflected.
Since in the analysis procedure we divide the interference spectrum by the non-interference background, we can remove the dependence on $P_0$, thus ${I_\text{LCI}}\propto \sqrt{R_s(d)}$. We can thus fit its experimental behavior with a function of the type $\sqrt{\mathrm{erfc}}$, and then recover the FWHM from the derivative of its square power. The experimental data and corresponding erfc fit function are reported in Fig. \ref{fig:knifeedge}. It allows us to estimate the idler beam diameter as $(3.3\pm0.5)$ mm.

\section{Simulation of the LCI roll-off curve}
\begin{figure}[!ht]
\includegraphics[scale=0.5]{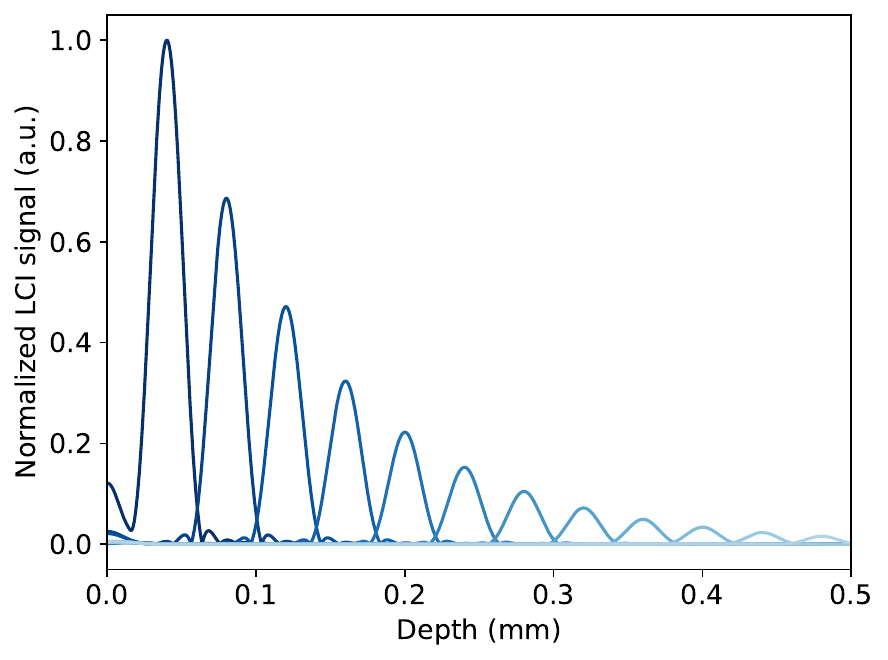}
\centering
\caption{\textit{Simulated rolloff curve for the first crystal design, assuming a spectrometer resolution equal to the experimental one $\delta \lambda=0.6$ $\mathrm{nm}$}.}
\label{fig:simulrolloff}
\end{figure}

To justify the experimental value of axial resolution and verify if the depth range is determined by the spectrometer's resolution, {we exploit the simulation of the LCI spectra.} 
{In the simulation, the spectrometer's response function is approximated by a Lorentzian function with the bandwidth related to the spectral resolution $\delta \lambda$, and $S_0(\omega_s)$ is taken from the experimentally measured non-interference background. For simplicity, $\rho(\omega_s)$ is neglected.} 
The simulated roll-off curve for the first crystal design, with poling period $\Lambda_1 = 12.3-14.0$ $\mathrm{\upmu m}$, is reported in Fig. \ref{fig:simulrolloff}. It exhibits axial resolution equal to $24$ $\mathrm{\upmu m}$, comparable to the experimental one of $29$ $\mathrm{\upmu m}$, and similar dependence of the LCI signal with the distance. Indeed, it decreases by 10 dB at $280$ $\mathrm{\upmu m}$ from the zero-path difference position, in good agreement with the $270$ $\mathrm{\upmu m}$ depth range found experimentally.

\section{Noise characterization}
\begin{figure}[!ht]
\includegraphics[scale=0.5]{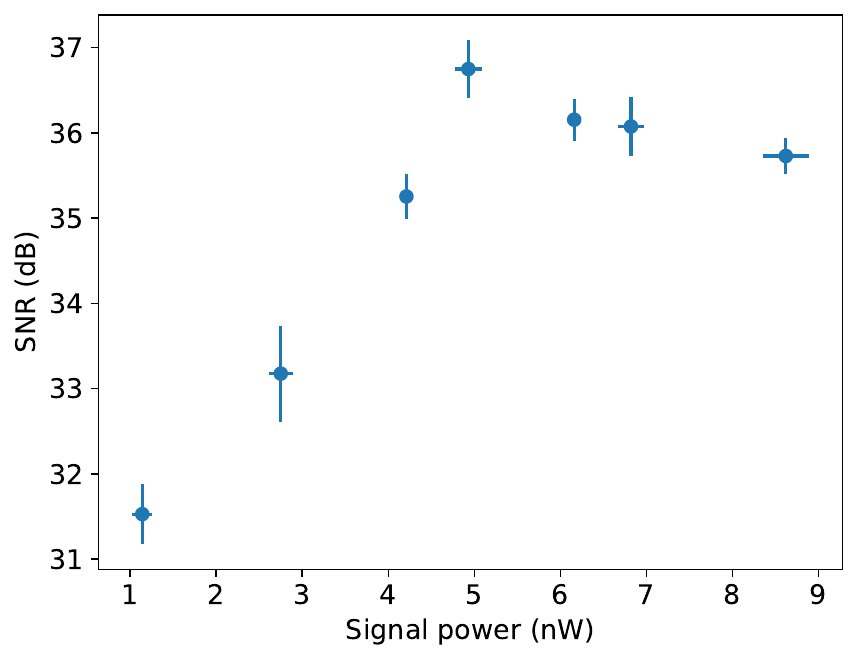}
\centering
\caption{\textit{The SNR dependence on the detected signal power. The uncertainties are given by the standard deviations of the SNR value and signal power, respectively.}}
\label{fig:noise}
\end{figure}

To identify the dominant source of noise, we experimentally reconstruct the SNR dependence on the detected signal power. We measure it by using a simple mirror at a fixed position as the sample, decreasing the detected signal power by rotating a continuously variable ND filter placed in front of the fiber before the spectrometer. The result is shown in Fig. \ref{fig:noise}. 

The SNR increases with the power for low signal powers, then saturates above $5$ $\mathrm{nW}$. This behavior is compatible with a change in the dominating noise, since the LCI signal is consistent with the expected linear behavior. In the operative region, i.e. high signal powers, the saturation of SNR suggests that the dominant noise is relative-intensity noise (RIN). Because of the linear scaling of both noise and LCI signal, we indeed expect their ratio to be constant, making SNR independent on the detected signal power. 

\section{Characterization of second stage amplification}
We estimate the ratio between signal power after the second OPA and idler power after the first OPA for several pump powers, all well above the standard operative value at $\approx 1$ $\mathrm{mW}$ to be able to catch the weak signal spectrum just after the first pass. The result is shown in Fig. \ref{fig:amplif}. As specified in the main text, we expect the ratio between detected signal power and probing idler power to scale like
\begin{equation}
    {\frac{P_\text{signal,2}}{P_\text{idler,1}}}\approx\frac{1}{2}\frac{\sinh^2(2r)}{\sinh^2(r)}\frac{\omega_s}{\omega_i}\approx\frac{1}{2}e^{2r}\frac{\omega_s}{\omega_i},
    \label{eq:exp_ampl_high}
\end{equation}
with $r\gg1$ the parametric gain. However, Eq. \ref{eq:exp_ampl_high} is only valid if both OPA processes are in the undepleted pump region. In the high-gain regime, the OPA process starts saturating at high pump energies, because of competitive nonlinear effects (i.e. second-harmonic generation and sum-frequency generation) increasingly depleting the pump \cite{pump_depletion}. Since the second stage OPA process is seeded by light from the first one, we also expect it to reach the pump depletion region quicker than the first stage OPA. As a consequence, we find that the ratio in Eq. \ref{eq:exp_ampl_high} decreases with the pump power. This is indeed what we experimentally observe in Fig. \ref{fig:amplif}, with the amplification factor decreasing almost linearly from circa $208$ to $115$. Nonetheless, all these values are an order of magnitude higher than what we can observe in the low-gain regime under the same conditions. Additionally, all the measured values are smaller or equal to the optimal amplification, given by the maximum of Eq. \ref{eq:exp_ampl_high} for undepleted pump. Therefore, we can fairly state $208$ is the best available estimation for it.    

\begin{figure}[!ht]
\includegraphics[scale=0.5]{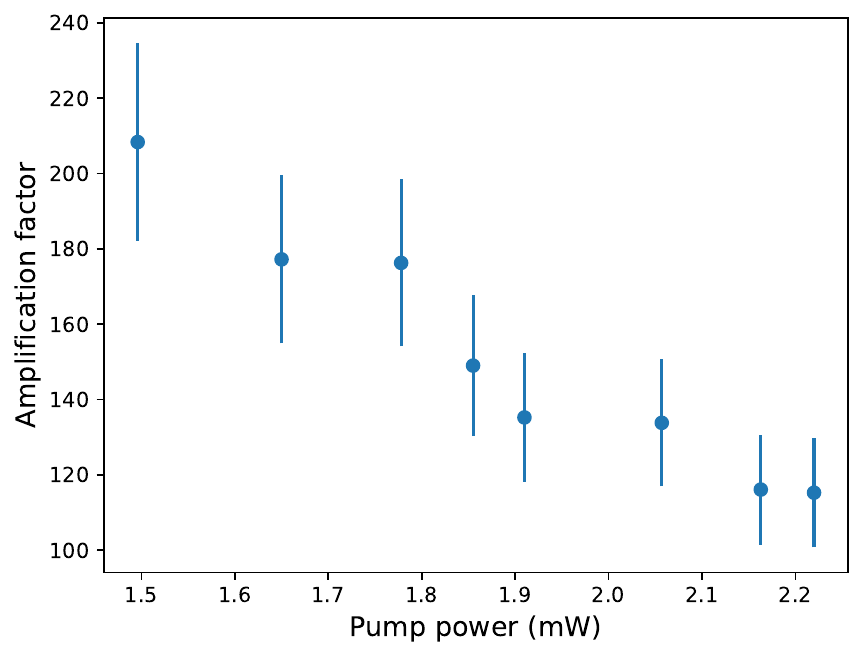}
\centering
\caption{\textit{The dependence of the amplification factor, defined as the ratio between the idler power after the first OPA and the signal power after the second one, as a function of the pump power. The uncertainties are obtained by propagating the standard deviations of the signal powers measured before and after the second OPA.}}
\label{fig:amplif}
\end{figure}

\bibliography{sample}

\begin{thebibliography}{10}
\newcommand{\enquote}[1]{``#1''}

\bibitem{LCIandOCT}
L.~J. Cusato, S.~Cerrotta, J.~R. Torga, and E.~N. Morel, \enquote{Extending low-coherence interferometry dynamic range using heterodyne detection,} {\protect\JournalTitle{Optics and Lasers in Engineering}} \textbf{131}, 106106 (2020).

\bibitem{drexler2008optical}
W.~Drexler and J.~G. Fujimoto, \emph{Optical coherence tomography: technology and applications} (Springer Science \& Business Media, 2008).

\bibitem{Su_ceramics_1}
R.~Su, M.~Kirillin, E.~W. Chang, \emph{et~al.}, \enquote{Perspectives of mid-infrared optical coherence tomography for inspection and micrometrology of industrial ceramics,} {\protect\JournalTitle{Opt. Express}} \textbf{22}, 15804--15819 (2014).

\bibitem{SiC_OCT}
M.~D. Duncan, M.~Bashkansky, and J.~Reintjes, \enquote{Subsurface defect detection in materials using optical coherence tomography,} {\protect\JournalTitle{Opt. Express}} \textbf{2}, 540--545 (1998).

\bibitem{Zorinpyroarray}
I.~Zorin, R.~Su, A.~Prylepa, \emph{et~al.}, \enquote{Mid-infrared fourier-domain optical coherence tomography with a pyroelectric linear array,} {\protect\JournalTitle{Optics Express}} \textbf{26}, 33428--33439 (2018).

\bibitem{FirstMIROCT}
C.~S. Colley, J.~C. Hebden, D.~T. Delpy, \emph{et~al.}, \enquote{Mid-infrared optical coherence tomography,} {\protect\JournalTitle{Review of Scientific Instruments}} \textbf{78}, 123108 (2007).

\bibitem{Israelsen_upconversion}
N.~M. Israelsen, C.~R. Petersen, A.~Barh, \emph{et~al.}, \enquote{Real-time high-resolution mid-infrared optical coherence tomography,} {\protect\JournalTitle{Light: Science \& Applications}} \textbf{8}, 11 (2019).

\bibitem{yagi_upconversion}
S.~Yagi, T.~Nakamura, K.~Hashimoto, \emph{et~al.}, \enquote{Mid-infrared optical coherence tomography with mhz axial line rate for real-time non-destructive testing,} {\protect\JournalTitle{APL Photonics}} \textbf{9}, 051301 (2024).

\bibitem{ChekhovaSU11}
M.~V. Chekhova and Z.~Y. Ou, \enquote{Nonlinear interferometers in quantum optics,} {\protect\JournalTitle{Adv. Opt. Photon.}} \textbf{8}, 104--155 (2016).

\bibitem{TashimaQPM}
T.~Tashima, Y.~Mukai, M.~Arahata, \emph{et~al.}, \enquote{Ultra-broadband quantum infrared spectroscopy,} {\protect\JournalTitle{Optica}} \textbf{11}, 81--87 (2024).

\bibitem{VanselowQPM}
A.~Vanselow, P.~Kaufmann, H.~M. Chrzanowski, and S.~Ramelow, \enquote{Ultra-broadband spdc for spectrally far separated photon pairs,} {\protect\JournalTitle{Opt. Lett.}} \textbf{44}, 4638--4641 (2019).

\bibitem{Paterova_first_LCI_und}
A.~V. Paterova, H.~Yang, C.~An, \emph{et~al.}, \enquote{Tunable optical coherence tomography in the infrared range using visible photons,} {\protect\JournalTitle{Quantum Science and Technology}} \textbf{3}, 025008 (2018).

\bibitem{Vanselow:20}
A.~Vanselow, P.~Kaufmann, I.~Zorin, \emph{et~al.}, \enquote{Frequency-domain optical coherence tomography with undetected mid-infrared photons,} {\protect\JournalTitle{Optica}} \textbf{7}, 1729--1736 (2020).

\bibitem{Ishakov}
T.~S. Iskhakov, A.~M. P\'{e}rez, K.~Y. Spasibko, \emph{et~al.}, \enquote{Superbunched bright squeezed vacuum state,} {\protect\JournalTitle{Opt. Lett.}} \textbf{37}, 1919--1921 (2012).

\bibitem{mWtwinbeams}
Y.~Eto, M.~Nuriya, and H.~Kano, \enquote{Quantum frequency-resolved optical gating measurement for ultranarrow temporal correlation of twin beams,} {\protect\JournalTitle{Optica Quantum}} \textbf{2}, 468--474 (2024).

\bibitem{Horoshko2020}
D.~Horoshko, M.~Kolobov, F.~Gumpert, \emph{et~al.}, \enquote{Nonlinear {Mach--Zehnder} interferometer with ultrabroadband squeezed light,} {\protect\JournalTitle{J. Mod. Opt.}} \textbf{67}, 41--48 (2020).

\bibitem{Machado}
G.~J. Machado, G.~Frascella, J.~P. Torres, and M.~V. Chekhova, \enquote{Optical coherence tomography with a nonlinear interferometer in the high parametric gain regime,} {\protect\JournalTitle{Applied Physics Letters}} \textbf{117}, 094002 (2020).

\bibitem{Hashimotohighgainvis}
K.~Hashimoto, D.~B. Horoshko, and M.~V. Chekhova, \enquote{Broadband spectroscopy and interferometry with undetected photons at strong parametric amplification,} {\protect\JournalTitle{Advanced Quantum Technologies}} \textbf{8}, 2300299 (2025).

\bibitem{Chekhovaapktp}
M.~V. Chekhova, S.~Germanskiy, D.~B. Horoshko, \emph{et~al.}, \enquote{Broadband bright twin beams and their upconversion,} {\protect\JournalTitle{Opt. Lett.}} \textbf{43}, 375--378 (2018).

\bibitem{HoroshkoapKTP}
D.~B. Horoshko and M.~I. Kolobov, \enquote{Generation of monocycle squeezed light in chirped quasi-phase-matched nonlinear crystals,} {\protect\JournalTitle{Phys. Rev. A}} \textbf{95}, 033837 (2017).

\bibitem{Hashimoto2024}
K.~Hashimoto, D.~B. Horoshko, M.~I. Kolobov, \emph{et~al.}, \enquote{Fourier-transform infrared spectroscopy with undetected photons from high-gain spontaneous parametric down-conversion,} {\protect\JournalTitle{Communications Physics}} \textbf{7}, 217 (2024).

\bibitem{rolloff}
A.~G. Podoleanu, \enquote{Optical coherence tomography,} {\protect\JournalTitle{Journal of Microscopy}} \textbf{247}, 209--219 (2012). Epub 2012 Jun 18.

\bibitem{Takeuchi_rolloffpulsed}
J.~Kaur, Y.~Mukai, R.~Okamoto, and S.~Takeuchi, \enquote{Spectral domain nonlinear quantum interferometry with pulsed laser excitation,} {\protect\JournalTitle{Phys. Rev. A}} \textbf{108}, 063714 (2023).

\bibitem{solarcellthick}
I.~Lombardero, M.~Ochoa, N.~Miyashita, \emph{et~al.}, \enquote{Theoretical and experimental assessment of thinned germanium substrates for iii–v multijunction solar cells,} {\protect\JournalTitle{Progress in Photovoltaics: Research and Applications}} \textbf{28}, 1097--1106 (2020).

\bibitem{photonnumbercorr}
M.~Bondani, A.~Allevi, G.~Zambra, \emph{et~al.}, \enquote{Sub-shot-noise photon-number correlation in a mesoscopic twin beam of light,} {\protect\JournalTitle{Phys. Rev. A}} \textbf{76}, 013833 (2007).

\bibitem{Hilbert}
M.~Wojtkowski, V.~J. Srinivasan, T.~H. Ko, \emph{et~al.}, \enquote{Ultrahigh-resolution, high-speed, {Fourier} domain optical coherence tomography and methods for dispersion compensation,} {\protect\JournalTitle{Opt. Express}} \textbf{12}, 2404--2422 (2004).

\bibitem{proHanning}
D.~Hillmann, T.~Bonin, C.~L\"{u}hrs, \emph{et~al.}, \enquote{Common approach for compensation of axial motion artifacts in swept-source {OCT} and dispersion in {Fourier-domain OCT},} {\protect\JournalTitle{Opt. Express}} \textbf{20}, 6761--6776 (2012).

\bibitem{Hanninglimit}
C.~B. Walker, A.~Wisniowiecki, J.~C. Tang, \emph{et~al.}, \enquote{Multi-window approach enables two-fold improvement in {OCT} axial resolution with strong side-lobe suppression and improved phase sensitivity,} {\protect\JournalTitle{Biomed Opt Express}} \textbf{14}, 6301--6316 (2023).

\bibitem{kalkman}
J.~Kalkman, \enquote{Fourier-domain optical coherence tomography signal analysis and numerical modeling,} {\protect\JournalTitle{International Journal of Optics}} \textbf{2017}, 9586067 (2017).

\bibitem{pump_depletion}
J.~Fl\'{o}rez, J.~S. Lundeen, and M.~V. Chekhova, \enquote{Pump depletion in parametric down-conversion with low pump energies,} {\protect\JournalTitle{Opt. Lett.}} \textbf{45}, 4264--4267 (2020).

\end{thebibliography}

\end{document}